\documentclass[aps,showpacs]{revtex4}
\usepackage{amsmath, amssymb,color,graphicx,subfigure}
\usepackage{float}
\begin{document}
\newcommand{\hh}{\hspace{0.2in}}
\newcommand{\vv}{\vspace{0.2in}}
\newcommand{\mbb}{\mathbf}
\newcommand{\ex}{\texttt{e}}
\newcommand{\mc}{\mathcal}
\title{Anisotropic plasmon-coupling dimerization of a pair of spherical electron gases}

\author{Godfrey Gumbs$^{1,2}$, Andrii Iurov$^{1}$, Antonios Balassis,$^{3}$  Danhong Huang$^{4}$}
\affiliation{$^{1}$Department of Physics and Astronomy,  Hunter College  of the \\
 City University of New York, 695 Park Avenue, New York, NY 10065, USA  \\
$^{2}$ Donostia International Physics Center (DIPC),
P de Manuel Lardizabal, 4, 20018 San Sebastian, Basque Country, Spain \\
$^{3}$Physics Department, Fordham University
441 East Fordham Road, Bronx, NY 10458, USA\\
$^{4}$Air Force Research Laboratory, Space Vehicles Directorate
Kirtland Air Force Base, NM 87117, USA}

\date{\today}

\begin{abstract}
We have discovered a novel feature in the plasmon excitations for a pair of Coulomb-coupled
non-concentric spherical  two-dimensional electron gases (S2DEGs). Our results show that
the plasmon excitations for such pairs depend on the orientation with respect to the
external electromagnetic probe field. The origin of this anisotropy of the inter-sphere
Coulomb interaction is due to the directional asymmetry of the electrostatic coupling
of electrons in excited states  which depend on both the angular momentum quantum number
$L$ and its    projection $M$  on the axis of quantization taken as the probe ${\bf E}$-field
direction. We demonstrate the anisotropic inter-sphere Coulomb coupling in 
space and present semi-analytic results in the random-phase approximation both perpendicular and parallel
to the axis of quantization. For the incidence of light with a finite orbital or spin angular momentum, 
the magnetic field generated from an induced oscillating electric dipole on one sphere can couple to an
induced magnetic dipole on another sphere in a way depending on the direction parallel or perpendicular
to the probe ${\bf E}$ field. Such an effect from the plasmon spatial correlation is expected to be 
experimentally observable by employing circularly-polarized light or a helical light beam for incidence.
The S2DEG serves as a simple model for fullerenes as well as metallic dimers, when the energy bands are 
far apart.
\end{abstract}


\maketitle

\section{Introduction}
\label{sec1}

Recent calculations on the plasma excitations of a spherical two-dimensional
electron gas (S2DEG) have yielded some interesting behaviors
as functions of the angular momentum quantum number $L$ and the
radius $R$ of the shell \,\cite{1,2,3}. In those model calculations,
the electron gas is assumed to be confined to an infinitesimally
thin shell which is embedded in a medium with background
dielectric constant $\epsilon_b$. The plasma excitation
frequencies were shown not to depend on the projection
$M$ of angular momentum $L$ on the axis of quantization. This degeneracy is
expected due to retained rotational symmetry and there is no
energy dispersion which arises in the case for the cylindrical nanotube\,\cite{4,5}.
Interest in plasmon excitations in fullerenes dates back to
the work by \"Ostling, et al.\,\cite{Apell} who used a spherical
shell model to examine the experimental data for plasma resonances
in C$_{60}$\,\cite{expt1,expt2}.The model for the plasmons described
in Ref.\,[\onlinecite{Apell}] assumes that the buckyball is doped and
its active modes are attributed to either dipole or monopole-like excitations. Since our
system is neutral, only dipole-like plasmon modes will exist.
\medskip

The interest in the S2DEG has been generated by the observation that
fullerenes\,\cite{O1,O1b,O2,O3,O4,O5,O6} span an entire family from ''buckybabies''  with
thirty-two carbon atoms and radius $0.35$\,nm to very large
fullerenes with four thousand, eight hundred and sixty atoms
and radius $3.141$\,nm\,\cite{O1,O1b,O2,O3,O4}. These molecules can be
modified into other molecular configurations which make them very versatile.
It is such an adaptability that gives them enormous practical applications
in materials science, electronics and nanotechnology.
\medskip

The model we employ, which consists of a S2DEG confined to the surface, allows
us to investigate the electronic properties, related to their
spherical shape and the lattice structure\,\cite{Lett1,Lett2,SSC},
and to neglect their radial motion\,\cite{SSC5} at the same time. These electronic properties include the
collective plasma excitations, electron energy loss spectra
as well as the thermoelectric properties of fullerenes.
This is conceptually similar to the electron gas model for
a carbon nanotube, which has been studied extensively\,\cite{4,18,20}.
The only variables in our calculations are the
radius, the separation between two displayed shells and the number of free electrons on each shell.
The plasma formula contains these parameters, as well as the orbital angular momentum,
and we vary them to examine how the plasmon frequency depends on them.
\medskip

\begin{figure}[ht!]
\centering
\includegraphics[width=0.35\textwidth]{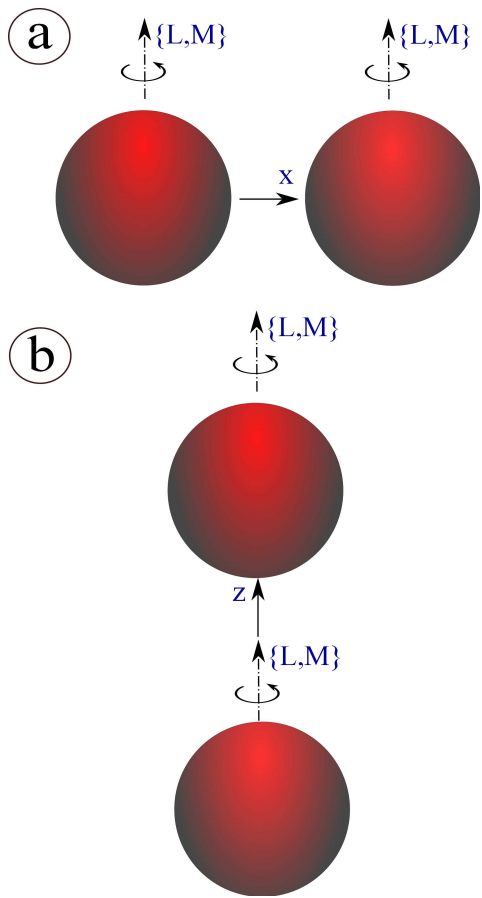}
\caption{(Color online) Schematic illustration of
a pair of displaced buckyballs. The axis of quantization is
along the $z$ direction with angular momentum quantum
number $L$ and component $M$. In (a), one buckyball
has its center at the origin and the other is centered on
the $x$-axis. In (b), one buckyball
has its center at the origin while the other is centered on
the $z$-axis.}
\label{FIG:1}
\end{figure}

The authors of Ref.\,[\onlinecite{SSC-1996}]
attempted to obtain the plasma excitations for a pair of displaced S2DEGs whose centers
do not coincide. However, the formalism by Rotkin and Suris\,\cite{SSC-1996}
is incomplete since these authors did not include the full Coulomb coupling
between the two S2DEGs in their Eqs.\,(3) and (4). Consequently,
the Eq.\,(8) in their paper for plasma excitations has missed the Coulomb
matrix elements coupling all the possible angular momenta on the
two spheres, meaning that we cannot use the angular momentum to label
plasma excitations, or $L$ is no longer a good quantum number for the system considered,
as they have done in their Eq.\,(9).
Another interesting subtlety which arises from the Coulomb dimer
is worth commenting on at this point. Suppose
we chose the axis of quantization for angular momentum to be
along the $z$ direction. Let us assume that one of the S2DEGs has its
center at the origin as shown in Fig.\,\ref{FIG:1}. We then have a
choice of placing the second S2DEG with its center on the $z$-axis or
on the $x$-axis. From a mathematical point of view, the Coulomb
matrix elements involved have values which depend on the spatial
coordinates. From a physical point of view, once the axis of
quantization is chosen, the spherical symmetry in the Coulomb dimer
is broken, resulting in a dependence of location for the second S2DEG.
Similar calculations of the polarization functions proved the 
existence of strongly localized image states near the surface of a buckyball \,\cite{ours}.

\medskip

\begin{figure}[ht!]
\centering
\includegraphics[width=0.2\textwidth]{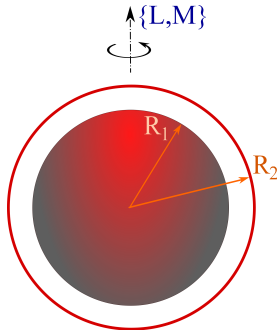}
\caption{(Color online) Schematic illustration of
a pair of concentric shells with inner radius $R_1$
and outer radius $R_2$. Here, $L$ and $M$ are
angular momentum quantum number and component for spheres.}
\label{FIG:1+}
\end{figure}

In this paper, we  are particularly interested in calculating
the plasma excitations of two Coulomb coupled S2DEGs. These may model
either a double shell onion-type buckyball as illustrated in
Fig.\,\ref{FIG:1+} or a pair of  non-overlapping buckyballs
whose centers have a finite separation between them (called Coulomb dimer), as depicted in
Fig.\,\ref{FIG:1}. Each S2DEG may be polarized by external
electromagnetic fields. However, the S2DEG is only
polarized for finite angular momentum quantum number $L\neq 0$. Additionally, $L$ is still a good
quantum number for labeling the plasma excitations on concentric shells.
Here, for simplicity, $L=0$ corresponds to a non-circularly-polarized probe field,
while $L=1$ is associated with a circularly-polarized probe field.
Moreover, the higher angular momentum with $L>1$ can be achieved by a special
light beam, e.g., a helical light beam.
However, when two S2DEGs have their centers well separated so
that there is no overlap of their charge distributions, the breaking of
the spherical symmetry leads to significant differences with the
double-wall buckyball, as we now describe.
\medskip

When the two shells are concentric, the Coulomb
interaction between them enters the plasmon mode equation through its
dependence on the angular momentum quantum number $L$ as well as the
radius of each shell. This type of Coulomb coupling only leads to a renormalization
of each of the two plasmon modes which exist on each shell independent
of the S2DEG on the other. The resulting coupled modes are
in-phase symmetric and out-of-phase antisymmetric charge-density
oscillations. However, as far as the Coulomb dimer is considered,
the polarization functions for all values of $L$
on each sphere are coupled to each other. The inter-sphere Coulomb matrix element
depends on both $L$ and its projection $M$ on the axis of quantization.
Therefore, in principle, the plasmon mode equation is given in terms of a
determinant of infinite dimension. But, the corresponding
matrix may be divided into diagonal sub-matrices corresponding to
$L=1,\,2,\,3,\,\cdots$ and consisting of $2(2L+1)\times 2(2L+1)$ elements
whose Coulomb interactions depend on $L$ and $(2L+1)$ values
of $M$ for each of the two shells.
The off-diagonal sub-matrices, on the other hand,
involve Coulomb matrix elements which depend on pairs of different
angular momenta, $L$ and $L^\prime$, arising from each sphere. These off-diagonal
Coulomb terms are generally smaller than their diagonal
counterparts and so may be formally treated as perturbations. Consequently, in
the lowest-order approximation, the $L=1$ mode is split
by Coulomb interactions depending on $M=0,\,\pm 1$  on each
sphere, leading to the occurrence of three symmetric and three antisymmetric
plasmon modes. Additionally, we derive semi-analytic expressions for these Coulomb
matrix elements for large separations. The Coulomb interaction described by the
diagonal sub-matrices directly lead to the spatial correlation between plasmons on
two spheres (or simply called the Coulomb dimerization for short).
\medskip

In Sec.\,\ref{sec2}, we will first formulate the method for
calculating the plasmon equations on a pair of non-overlapping S2DEGs.
This is based on the random-phase approximation (RPA) in
evaluating the induced density fluctuations for a weak external
perturbation. Section\ \ref{sec3} is devoted to a discussion
of our numerical results. Some concluding remarks are given in
Sec.\,\ref{sec4}. Mathematical details of our calculations are
provided in Appendices.

\section{General Formulation of the Problem}
\label{sec2}

\subsection{Plasmon-Mode Equation for a Double-Walled Nano-Sphere}

We first use linear-response theory to derive the plasma
mode equation for a S2DEG on a pair of concentric shells
with inner radius $R_1$ and outer radius $R_2$.
If an electron is confined on the surface of a sphere
of radius $R$, the eigenfunctions and eigenenergies are

\begin{mathletters}
\label{gae1}
\begin{equation}
<{\bf r}|\alpha>\equiv\psi_{\alpha}(r,\theta,\varphi)=\frac{{\cal R}(r)}{R}\,
Y_{\ell,m}(\Omega)\ ,\ \ \ \ {\cal R}^2(r)=\delta(r-R)\ ,
\label{gae1a}
\end{equation}
\begin{equation}
\varepsilon_{\alpha}=\frac{\ell(\ell+1)\hbar^2}{2\mu^{\ast}R^2}
\label{gae1b}
\end{equation}
\end{mathletters}
with $R=R_1$ or $R_2$, $Y_{\ell,m}(\Omega)\equiv Y_{\ell,m}(\theta,\,\phi)$ being the spherical harmonic function,
$\mu^\ast$ the effective mass of electrons, $\displaystyle{\alpha=\{\ell,\,m\}}$, $\displaystyle{|m|\leq\ell}$ and
$\ell=0,\,1,\,2,\,\cdots$. The density matrix is given by

\begin{eqnarray}
&&<\ell m|\hat{\rho}_1(\omega)|\ell^{\prime}m^{\prime}>=-2e\sum\limits_{i=1}^2\,
\frac{f_0(\varepsilon_{\ell}^i)-f_0(\varepsilon_{\ell^{\prime}}^i)}
{\hbar\omega+\varepsilon_{\ell}^i-\varepsilon_{\ell^{\prime}}^i}
\nonumber\\
&&\times\sum\limits_{L,M}\,{\cal F}_{L,\,M}(R_i,\,\omega)
\int\,d\Omega\,Y_{\ell,\,m}^\ast(\Omega)
Y_{L,\,M}(\Omega)\,Y_{\ell^{\prime},\,m^{\prime}}(\Omega)\ ,
\label{gae22}
\end{eqnarray}
where, after some calculation, we obtain from the Poisson equation with respect to the density fluctuation

\begin{equation}
\frac{1}{r^2}\,\frac{d}{dr}\left[r^2\,\frac{d{\cal F}_{L,\,M}(r,\,\omega)}{dr}\right]-\frac{L(L+1)}{r^2}\,
{\cal F}_{L,\,M}(r,\,\omega)
=\sum\limits_{i=1}^2\,{\cal A}_{L,\,M}^{(i)}(\omega)\,\delta(r-R_i)
\label{gae23}
\end{equation}
with $\omega$ being the external-field frequency and

\begin{eqnarray}
{\cal A}_{L,\,M}^{(i)}(\omega) = -{\cal F}_{L,\,M}(R_i,\,\omega)\,\frac{e^2}{\epsilon_sR_i^2}\,
\Pi_{L}^{(i)}(\omega)\ .
\ \label{gae24}
\end{eqnarray}
Here, $\epsilon_s\equiv 4\pi\epsilon_0\epsilon_b$,
$\epsilon_b$ is the uniform background dielectric
constant, the polarization function is given by

\begin{equation}
\Pi_{L}^{(i)}(\omega)=2\sum\limits_{\ell,\ell^\prime}\,
\frac{f_0(\varepsilon_{\ell}^{i})-f_0(\varepsilon_{\,\ell^\prime}^{i})}
{\hbar\omega+\varepsilon_{\ell}^{i}-\varepsilon_{\ell^\prime}^{i}}\,
(2\ell+1)(2\ell^\prime+1)
\left( \begin{matrix}
\ell & \ell^\prime & L\cr
0   & 0           & 0\cr
\end{matrix}\right)^2
\label{gae13}
\end{equation}
in terms of the Wigner 3j-symbol, the Fermi-Dirac distribution
function $f_0(\varepsilon_{\ell}^{i}) $ and the eigenenergies $\varepsilon_{\ell}^{i}$
are obtained from Eq.\,(\ref{gae1b})
by replacing $R$ with $R_i$.
\medskip

Now, ${\cal F}_{L,\,M}(r,\,\omega)$ is explicitly given by

\begin{equation}
r{\cal F}_{L,\,M}(r,\,\omega)=\left\{ \begin{matrix}
\overline{C}_1\,r^{L+1}\ , & r <R_1\cr
\overline{C}_3\,r^{L+1}+\overline{C}_4\,r^{-L}\ , & R_1\leq r\leq R_2\cr
\overline{C}_2\,r^{-L}\ , & r>R_2\cr
\end{matrix}\right.\ ,
\label{gae25}
\end{equation}
where $\overline{C}_1$, $\overline{C}_2$, $\overline{C}_3$
and $\overline{C}_4$ are some
constants to be determined from the continuity of
${\cal F}_{L,\,M}(r,\,\omega)$ at $r=R_1,\,R_2$ and the step-like change of
$\displaystyle{\frac{d[r{\cal F}_{L,\,M}(r,\,\omega)]}{dr}}$ at $r=R_1,\,R_2$. From these boundary conditions, we get

\begin{equation}
\left[\begin{matrix}
R_1^{L+1} & 0 & -R_1^{L+1} & -R_1^{-L}\cr
0 & -R_2^{-L} & R_2^{L+1} & R_2^{-L}\cr
-(L+1)R_1^{L} & 0 & (L+1)R_1^{L} & -LR_1^{-(L+1)}\cr
0 & -LR_2^{-(L+1)} & -(L+1)R_2^{L} & LR_2^{-(L+1)}\cr
\end{matrix}\right]
\left[\begin{matrix}
\overline{C}_1\cr
\overline{C}_2\cr
\overline{C}_3\cr
\overline{C}_4\cr
\end{matrix} \right]=\left[\begin{matrix}
0\cr 0\cr
R_1{\cal A}_{L,\,M}^{(1)}\cr
R_2{\cal A}_{L,\,M}^{(2)}\cr
\end{matrix}\right]\ ,
\label{gae26}
\end{equation}
which we have solved for $\overline{C}_1,\,\overline{C}_2,\,
\overline{C}_3$  and $\overline{C}_4$ and obtained

\begin{eqnarray}
\overline{C}_1&=&-\frac{R_1^{1-L}}{2L+1}\,A^{(1)}_{L,\,M}(\omega)
-\frac{R_2^{1-L}}{2L+1}\,A^{(2)}_{L,\,M}(\omega)\ ,
\nonumber\\
\overline{C}_2&=&-\frac{R_1^{L+2}}{2L+1}\,A^{(1)}_{L,\,M}(\omega)
-\frac{R_2^{L+2}}{2L+1}\,A^{(2)}_{L,\,M}(\omega)\ ,
\nonumber\\
\overline{C}_3&=&-\frac{R_2^{1-L}}{2L+1}\,A^{(2)}_{L,\,M}(\omega)\ ,
\nonumber\\
\overline{C}_4&=&-\frac{R_1^{L+2}}{2L+1}\,A^{(1)}_{L,\,M}(\omega)\ .
\label{gae27}
\end{eqnarray}
\medskip

Combining Eqs.\,(\ref{gae24}) and (\ref{gae25}) and making use of the results
for $\overline{C}_1$ and $\overline{C}_2$ in Eq.\,(\ref{gae27}), we obtain the
following pair of simultaneous equations

\begin{equation}
\left[\begin{matrix}
1-\displaystyle{\frac{e^2}{\epsilon_s(2L+1)R_1}\,\Pi_{L}^{(1)}(\omega)} &
-\displaystyle{\frac{e^2}{\epsilon_s(2L+1)R_1}\,\left(\frac{R_1}{R_2}\right)^{L+1}\Pi_{L}^{(2)}(\omega)} \cr
-\displaystyle{\frac{e^2}{\epsilon_s(2L+1)R_2}\left(\frac{R_1}{R_2}\right)^L\Pi_{L}^{(1)}(\omega)} &
1-\displaystyle{\frac{e^2}{\epsilon_s(2L+1)R_2}\,\Pi_{L}^{(2)}(\omega)}
\end{matrix}\right]
\left[\begin{matrix}
{\cal F}_{L,\,M}(R_1,\,\omega)\cr
\\
{\cal F}_{L,\,M}(R_2,\,\omega)
\end{matrix}\right]=0\ .
\label{gae28}
\end{equation}
Equation (\ref{gae28})  has non-trivial solutions
for ${\cal F}_{L,\,M}(R_1,\,\omega)$ and ${\cal F}_{L,\,M}(R_2,\,\omega)$ only if the determinant
of the coefficient matrix is zero, i.e.,

\begin{eqnarray}
&&1-\frac{e^2}{\epsilon_s(2L+1)}
\left[\frac{\Pi_{L}^{(1)}(\omega)}{R_1}
+\frac{\Pi_{L}^{(2)}(\omega)}{R_2}\right]
\nonumber\\
&&+\left[\frac{e^2}{\epsilon_s(2L+1)}\right]^2\,
\left[1-\left(\frac{R_1}{R_2}\right)^{2L+1}\right]\,
\frac{\Pi_{L}^{(1)}(\omega)\Pi_{L}^{(2)}(\omega)}{R_1R_2}=0\ .
\label{gae29}
\end{eqnarray}

\subsection{Plasmon Modes for a Pair of Displaced Spherical Shells}

We now turn our attention to a system of two spherical shells with their
centers on the $x$ axis. The center of one of the spheres is
at $x=0$ with radius $R_1$ whereas the other sphere is centered
at $x=a$ and its radius is $R_2$. We assume that the inequality
$a>R_1+R_2$ is satisfied to ensure no overlapping of charge distributions. In the absence of electron tunneling
between the shells, the wave function for an electron on the
$j$-th shell ($j=1,2$) is given by

\begin{equation}
<{\bf r}\mid j\nu>\equiv\Psi_{j,\ell m}\left({\bf r}-(j-1)a\hat{\bf e}_x\right)=\frac{{\cal R}(r_j^\prime)}{R_j}\,Y_{\ell,\,m}(\Omega)
\label{eq1.37}
\end{equation}
with $\nu=\{\ell,\,m\}$ and ${\cal R}^2(r_j^\prime)=\delta(r_j^\prime-R_j)$. The
energy spectrum has the form of Eq.\,(\ref{gae1b})

\begin{equation}
\varepsilon_{j,\,\nu}=\frac{\ell(\ell+1)\hbar^2}{2\mu^{\ast}R_j^2}\ .
\label{eq1.38}
\end{equation}
The equation of motion for the density matrix operator is

\begin{equation}\label{eq1.39}
\imath\hbar\,\frac{\partial\hat{\varrho}}{\partial t}=\left[\hat{\cal H},\,\hat{\varrho}\right]_{-}\ ,
\end{equation}
where $\hat{\cal H}=\hat{\cal H}_0-e\Phi$ is the Hamiltonian of the electron
on the surface of the sphere, $\hat{\cal H}_0$ is the free-electron
Hamiltonian and $\Phi$ is the induced potential. The potential
$\Phi$ satisfies Poisson's equation

\begin{equation}
\nabla^2\Phi({\bf r},\,\omega)=\frac{4\pi e}{\epsilon_s}\,
\delta n({\bf r},\,\omega)\ .
\label{eq1.40}
\end{equation}
Additionally, $\delta n({\bf r},\,\omega)$ is the induced electron-density fluctuation.
We employ linear response theory (see Appendix\ \ref{ap1}) to calculate the
induced density fluctuation as

\begin{equation}
\delta n({\bf r},\,\omega)=\sum\limits_{j,\,j\prime}\,
\sum\limits_{\nu,\,\nu^\prime}\,<{\bf r}\mid j\nu>
<j\nu\mid\hat{\varrho}_{1}({\bf r},\,\omega)\mid j^{\prime}\nu^\prime>
<j^{\prime}\nu^{\prime}\mid{\bf r}>\ ,
\label{eq1.41}
\end{equation}
where

\begin{equation}
<j\nu\mid\hat{\varrho}_1({\bf r},\,\omega)\mid
j^{\prime}\nu^\prime>=-2e\,\frac{f_0(\varepsilon_{j,\,\nu})-
f_0(\varepsilon_{j^{\prime},\nu^\prime})}{\hbar\omega
+\varepsilon_{j,\,\nu}-\varepsilon_{j^{\prime},\nu^\prime}}
<j\nu\mid\Phi({\bf r},\,\omega)\mid
j^{\prime}\nu^\prime>\ ,
\label{eq1.42}
\end{equation}
and we express the induced potential as $\displaystyle{\Phi({\bf
r},\,\omega)=\frac{1}{\cal V}\sum\limits_{\bf q^\prime}\,\Phi({\bf q^\prime},\,\omega)\,
e^{i{\bf q^\prime}\cdot{\bf r}}}$ with ${\cal V}$ being the system volume. Then, Eq.\,(\ref{eq1.41}) becomes

\begin{eqnarray}
\delta n({\bf r},\,\omega)&=&-\frac{2e}{\cal V}\sum\limits_{j,\,j^\prime}
\sum\limits_{\nu,\,\nu^\prime}\,\frac{f_0(\varepsilon_{j,\,\nu})-
f_0(\varepsilon_{j^{\prime},\nu^\prime})}
{\hbar\omega+\varepsilon_{j,\,\nu}-\varepsilon_{j^{\prime},\nu^\prime}}
<{\bf r}\mid j\nu><j^\prime\nu^\prime\mid{\bf r}>\nonumber\\
&\times&\sum\limits_{{\bf q^\prime}}\Phi({\bf q^\prime},\,\omega)<j\nu\mid
e^{i{\bf q^\prime\cdot r}}\mid j^\prime\nu^\prime> \ ,
\label{eq1.43}
\end{eqnarray}
or by taking the Fourier transform with respect to ${\bf r}$

\begin{eqnarray}
\delta n({\bf q},\,\omega)&=&-\frac{2e}{\cal V}\sum\limits_{j,\,j^\prime}
\sum\limits_{\nu,\,\nu^\prime}\,\frac{f_0(\varepsilon_{j,\,\nu})-
f_0(\varepsilon_{j^{\prime},\nu^\prime})}
{\hbar\omega+\varepsilon_{j,\,\nu}-\varepsilon_{j^{\prime},\nu^\prime}}
<j^\prime\nu^\prime\mid e^{-i{\bf q}\cdot{\bf r}}\mid j\nu>
\nonumber\\
&\times&\sum\limits_{\bf q^\prime}\Phi({\bf q^\prime},\,\omega)
<j\nu\mid e^{i\bf q^\prime\cdot r}\mid j^\prime\nu^\prime>\ .
\label{eq1.44}
\end{eqnarray}
The matrix elements $<j\nu\mid e^{i\bf q\cdot r}\mid
j^\prime\nu^\prime>$ with wave functions $<{\bf r}\mid j\nu>$ given
by Eq.\,(\ref{eq1.37}) can be calculated based on the
expansion of a plane wave in spherical waves

\begin{equation}
e^{i\textbf{q}\cdot \textbf{r}\ }=4\pi\sum\limits_{L,\,M} \ \ i^L\,
j_L(qr)\,Y_{L,M}^\ast(\Omega_{\textbf{q}})
Y_{L,M}(\Omega)\ ,
\label{eq1.45}
\end{equation}
where $\Omega_{\textbf{q}}=\{\theta_{\textbf{q}},\,\phi_{\textbf{q}}\}$ in the ${\bf q}$-space and
$j_\ell(x)$ is a spherical Bessel function. The result is

\begin{eqnarray}
&& <j\nu\mid e^{i\bf q\cdot r}\mid j^\prime\nu^\prime>
=4\pi\,\delta_{jj^\prime}\,
e^{i(j-1)q_xa}
\nonumber\\
&\times&\sum\limits_{L,M}\,i^L\,j_L(qR_j)\,Y_{L,M}^\ast(\Omega_\textbf{q})
\int d\Omega\,Y^{\ast}_{\ell,m}(\Omega)
Y_{L,M}(\Omega)
Y_{\ell^{\prime},m^{\prime}}(\Omega)\ .
\label{eq1.46}
\end{eqnarray}

Substituting Eq.\,(\ref{eq1.46}) into Eq.\,(\ref{eq1.44}), we obtain after some
algebra

\begin{eqnarray}
&& \delta n({\bf q},\,\omega)=-(4\pi)^2\,\frac{2e}{\cal V}
\sum\limits_{\ell,\,m}
\sum\limits_{l^\prime,m^\prime}
\sum\limits_{j=1}^{2}\,
\frac{f_0(\varepsilon_{j,\ell})-f_0(\varepsilon_{j,\ell^{\,\prime}})}
{\hbar\omega+\varepsilon_{j,\ell}-\varepsilon_{j,\ell^{\,\prime}}}\,e^{-i(j-1)q_xa}
\nonumber\\
&\times&\sum\limits_{L,M}\,(-i)^L\,j_L(qR_j)\,
Y_{L,M}(\Omega_\textbf{q})
\int d\Omega\,Y^{\ast}_{\ell^\prime,m^\prime}(\Omega)
Y_{L,M}^\ast(\Omega)Y_{\ell,m}(\Omega)
\nonumber\\
&\times& \sum\limits_{q_x^\prime,\,{\bf q}^\prime_\perp}\,
e^{i(j-1)q_x^\prime a}\,
\Phi\left(q_x^\prime,{\bf q}^\prime_\perp,\,\omega\right)
\nonumber\\
&\times&\sum\limits_{L^\prime,M^\prime}\,i^{L^\prime}\,
j_{L^\prime}(q^\prime R_j)
Y_{L^\prime,M^\prime}^\ast(\Omega_{\textbf{q}^\prime})
\int d\Omega^\prime\,Y^{\ast}_{\ell,m}(\Omega^\prime)
Y_{L^{\prime},M^{\prime}}(\Omega^\prime)
Y_{\ell^{\prime},m^{\prime}}(\Omega^\prime)\ ,
\label{eq1.47}
\end{eqnarray}
where ${\bf q}^\prime_\perp=\{q_y^\prime,\,q_z^\prime\}$, or

\begin{eqnarray}
&& \delta n({\bf q},\,\omega)=-(4\pi)\,\frac{2e}{\cal V}\sum\limits_{j=1}^{2}\sum\limits_{L,M}
\sum\limits_{\ell,\,\ell^\prime}\,
\frac{f_0(\varepsilon_{j,\ell})-f_0(\varepsilon_{j,\ell^{\,\prime}})}
{\hbar\omega+\varepsilon_{j,\ell}-\varepsilon_{j,\ell^{\,\prime}}}\,
(2\ell+1)(2\ell^\prime+1)
\left( \begin{matrix}
\ell&\ell^\prime& L\cr
0 & 0 & 0\cr
\end{matrix}\right)^2
\nonumber\\
&\times&   e^{-i(j-1)q_xa}\,j_L(qR_j)\,
Y_{L,M}(\Omega_{\textbf{q}})
\nonumber\\
&\times& \sum\limits_{q_x^\prime,\,{\bf q}_\perp^\prime}\,
e^{i(j-1)q_x^\prime a}\,
\Phi\left(q_x^\prime,\,{\bf q}_\perp^\prime,\,\omega\right)\,
j_{L}(q^\prime R_j)\,
Y_{L,M}^\ast(\Omega_{\textbf{q}^\prime})\ .
\label{eq1.47+}
\end{eqnarray}
\medskip

Taking the Fourier transform of Eq.\,(\ref{eq1.40}), we have
$\displaystyle{\Phi({\bf q},\,\omega)=-\frac{4\pi e}{\epsilon_sq^2}\,\delta n({\bf q},\,\omega)}$. Using
this relation in Eq.\,(\ref{eq1.47}), we obtain

\begin{equation}
\delta n({\bf q},\omega)=\frac{(4\pi)^2e^2}{\epsilon_s}\sum\limits_{j,\,L,M}\,\Pi_{j,\,L}(\omega)\,
e^{-i(j-1)q_xa}\,j_L(q R_j)\,Y_{L,M}(\Omega_{\textbf{q}})\,U_{j,\,LM}(\omega)\ ,
\label{eq1.48}
\end{equation}
where $\Pi_{j,\,L}(\omega)$ is the density response function of
the $j$-th nano shell given by an expression similar to Eq.\,(\ref{gae13}) and

\begin{equation}
U_{j,\,LM}(\omega) = \frac{1}{\cal V}\sum\limits_{q_x,{\bf q}_\perp}\,e^{i(j-1)q_xa}\,
\frac{\delta n(q_x,\,{\bf q}_\perp,\,\omega)}{q_x^2+q_\perp^2}\,
j_L(qR_j)\,Y_{L,M}^\ast(\Omega_{\textbf{q}})\ .
\label{eq1.49}
\end{equation}
Substituting the expression for $\delta n({\bf q},\,\omega)$ given by Eq.\,(\ref{eq1.48}) into
Eq.\,(\ref{eq1.49}), we obtain

\begin{equation}
\sum\limits_{j^\prime=1}^{2}\sum\limits_{L^\prime=0}^{\infty}
\sum\limits_{M^\prime=-L^\prime}^{L^\prime}\,
\left[\delta_{jj^\prime}\,\delta_{LL^\prime}\,\delta_{MM^\prime}-
\Pi_{j^\prime,L^\prime}(\omega)\,
V_{LM,\,L^\prime M^\prime}(R_j,\,R_{j^\prime};\,a)\right]\,
U_{j^\prime,\,L^\prime M^\prime}(\omega)=0\ ,
\label{eq1.50}
\end{equation}
where the Coulomb-matrix elements are

\begin{eqnarray}
&&V_{LM,\,L^\prime M^\prime}(R_j,\,R_{j^\prime};\,a)
=\frac{2e^2}{\pi\epsilon_s}\int
\frac{d^3\textbf{q}}{q^2}\,
j_L(qR_j)j_{L^\prime}(qR_{j^\prime})\,
Y_{L,M}^\ast(\Omega_{\textbf{q}})Y_{L^\prime,M^\prime}(\Omega_{\textbf{q}})\,
e^{i(j-j^\prime)q_xa}
\nonumber\\
&=& \frac{e^2}{\epsilon_s(2L+1)R}\,\delta_{L,L^\prime}\,\delta_{M,M^\prime}
\  \  \  \mbox{when $j=j^\prime$ and $R_j=R_{j^\prime}=R$}\ .
\label{eq1.51}
\end{eqnarray}
As a result, we obtain explicitly, by setting $j=1,\,2$ in turn for each of
the two spheres,

\begin{eqnarray}
& & \left[1-\frac{e^2}{\epsilon_s(2L+1)R_j}\,
\Pi_{j,\,L}(\omega)\right]\,U_{j,\,LM}(\omega)
\nonumber\\
&-&\sum\limits_{j{\,^\prime}\neq j}\sum\limits_{L^\prime,M^\prime}\,
\Pi_{j^\prime,L^\prime}(\omega)\,
V_{LM,\,L^\prime M^\prime}(R_j,\,R_{j^\prime};\,a)\,
U_{j^\prime,\,L^\prime M^\prime}(\omega)=0\ .
\label{eq1.51b}
\end{eqnarray}
If the spheres are identical, then we
need only keep $j=1$, but still have to do the sum over $j^\prime=1,2$.
The set of linear equations in (\ref{eq1.51b}) has
nontrivial solutions provided that the determinant of the coefficient
matrix for $\{U_{j,\,LM}(\omega)\}$ is zero.  Consequently, plasmon
modes with different values of $L $ on the two shells can now be
coupled via the inter-sphere ($j\neq j^\prime$) Coulomb interaction. Since
$V_{1LM,\,2L^\prime M^\prime}(R_1,\,R_2;\,a)\to 0$ in the limit
$a\to\infty$, the coefficient matrix becomes diagonal when $a\gg R_1,R_2$ and the
plasmon-mode equation simply reduces to the result for isolated shell

\begin{equation}
\prod_{L}\,\left[\epsilon_{1,\,L}(\omega)\right]^{2L+1}\,
\left[\epsilon_{2,\,L}(\omega)\right]^{2L+1}=0\ ,
\label{eq1.54}
\end{equation}
where $\epsilon_{j,\,L}(\omega)$ is the dielectric function for
the $j$-th shell. The significance of equations  (\ref{eq1.51b})
for chosen $L,M$  is that they give explicitly the effect of
the Coulomb interaction on  each  shell through $\epsilon_{j,\,L}(\omega)$  as
well as the coupling  between the pair of shells through the
Coulomb matrix elements $V_{LM,\,L^\prime M^\prime}(R_j,\,R_{j^\prime};\,a)$, i.e., dimerization.
Additionally, the nature of this coupling may be characterized in the following
way when carrying out numerical calculations. For chosen $L$     and
$|M|\leq L$, there are $2(2L+1)\times 2(2L+1)$ elements in a block
sub-matrix which includes $(2L+1)$ elements along the diagonal, equal to
$\epsilon_{1,\,L}(\omega)$, and $(2L+1)$ diagonal elements
equal to $\epsilon_{2,\,L}(\omega)$. For example, if we
consider the coupling between
sub-matrices with angular momentum $L=1,\,2,\,3,\,\cdots,\,N$, then
the dimension of the matrix is $\displaystyle{2\sum\limits_{L=1}^N\,(2L+1)=2N^2+4N}$.
Specifically, if we use just the $L=1$ sub-matrix, we have
a $6\times 6$ matrix whose properties are discussed in the Appendix\ \ref{ap2}.

\section{Numerical Results}
\label{sec3}

  if the radius of the buckyball is small, we may neglect the bandwidth (compared to large
	energy level separation) due to electron hopping between neighboring lattice sites on the sphere
and simply use a spherical model of an  electron gas to describe its optical response.
In our numerical calculations, we chose $T=0$\,K and assume the number of occupied energy levels, $N_F$, is fixed.
Also, the electron effective mass $\mu^\ast=0.25\,m_e$ where $m_e$ is the free-electron mass. Additionally,
the frequency  in the polarization function is replaced by $\omega+i\gamma$ where we
chose the broadening parameter $\hbar\gamma=0.05$\,eV. The background
dielectric constant $\epsilon_b=2.4$ which is the same as for graphene.

\begin{figure}[ht!]
\centering
\includegraphics[width=0.28\textwidth]{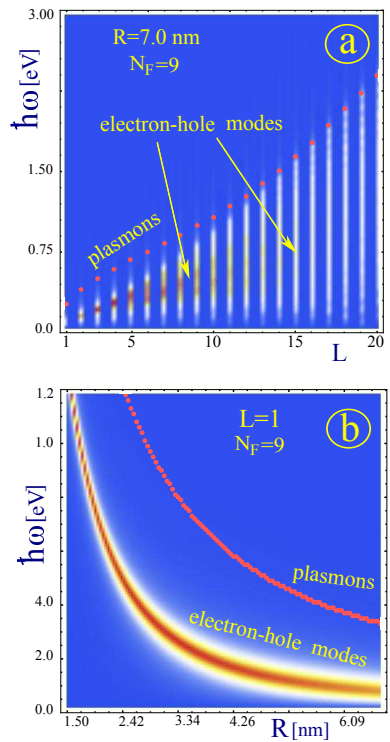}
\caption{(Color online) Density plots of frequencies $\omega$ for the plasmon
excitations and particle-hole modes on
a spherical shell as a function of $L$ (a) for $R=7$\,nm, $N_F=9$
and (b) as a function of its radius $R$  for $L=1$, $N_F=9$.}
\label{FIG:2}
\end{figure}
\medskip

Figure\ \ref{FIG:2}(a) shows the dependence of the plasmon excitation energy
as well as that for the single-particle modes on the angular momentum
quantum number $L$. Clearly, this dependence is similar to that for the 2DEG
when the plasma excitation energy is plotted as a function of wave number $q$.
In the long-wavelength limit, the plasmon frequency for the 2DEG
has a $\sqrt{q}$ dependence which resembles the variation
of frequency with $L$ in Fig.\,\ref{FIG:2}(a) at small angular momentum.
At large angular momentum, on the other hand, the plasmons are severely damped by the particle-hole
modes analogous to the Landau damping of plasmon excitations in a 2DEG
when $q$ is comparable with the Fermi wave vector $k_F$.
We observe that the highest intensity (largest values of ${\rm Im}\left[\Pi_L(\omega)\right]$)
for the electron-hole modes occur
at small $L$ values, i.e., $1\stackrel{<}{\sim} L\stackrel{<}{\sim} 3$.
\medskip

The corresponding region of high plasmon intensity lies close to the upper
boundary of single-particle excitations, as seen from Fig.\,\ref{FIG:2}(a).
Of course, these boundaries for electron-hole modes are determined by the chemical potential
of the S2DEG, i.e., the number of electrons, as well as the electron effective mass.
On the other hand, there is no analogy for the 2DEG with
Fig.\,\ref{FIG:2}(b) where we vary the radius of the S2DEG but keep
the number of electrons fixed. The plasmon mode frequency
decreases with increasing radius $R$ as $1/R^2$ for small $R$ values but $1/R$ for large
values of $R$. After the Fermi energy $E_F$ has been determined, one may label the topmost
occupied energy level by the angular momentum quantum number $\ell=\ell_F$,  while the next
empty level for active optical transition with $L=1$ may be labeled by $\ell^\prime=\ell_F+1$.
Physically, the Fermi energy $E_F$ cannot be fixed for discrete energy levels due to pinning of the
Fermi level, and it should be determined by the given total number of electrons, $N_e$, on the shell.
If the total number of occupied energy levels is $N_F$, we find $N_F=\sqrt{N_e/2}$ and the Fermi
energy is calculated through  $E_F=\ell_F(\ell_F+1)\hbar^2/(2\mu^\ast R^2)$ with $\ell_F=N_F-1$.
In addition, from the Fermi energy, one also gets the estimate  $(N_F-1)^2\approx 6.6\,[E_F({\rm
eV})]\,[R^2({\rm nm})]$ for $\ell_F\gg 1$, implying $E_F$ will depend on both $N_e$ and $R$ at the same time.
\medskip

\begin{figure}[ht!]
\centering
\includegraphics[width=0.28\textwidth]{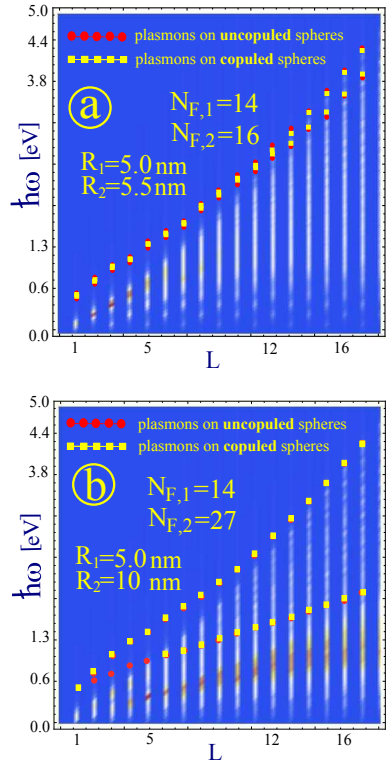}
\caption{(Color online) Density plot of frequency of the plasmon  excitations and particle-hole
modes {\em vs.} $L$ on two concentric S2DEGs for two various cases of close [in (a)] and
different [in (b)] radii of inner and outer spheres. Here, we chose $N_F=14$ for
$R_1=5.0$\,nm and $N_F=16$ for $R_2=5.5$\,nm in (a), while we choose $N_F=14$ for
$R_1=5.0$\,nm and $N_F=27$ for $R_2=10$\,nm in (b).}
\label{FIG:2+}
\end{figure}

We now turn to a description of our results in Fig.\,\ref{FIG:2+} for
the plasmon excitations of two concentric S2DEGs when the inner radius
is chosen as $R_1=5.0$\,nm and  the  outer radius $R_2=5.5$\,nm in (a) and
$R_2=10.0$\,nm in (b). Since the structure is spherically symmetric,
the plasma modes can still be labeled by the angular momentum
quantum number $L$. Both inner and outer shells have a single-particle
excitation spectrum which overlap when plotted as a function of $L$.
Additionally, each S2DEG gives rise to a plasmon branch which is renormalized
by the inter-shell Coulomb interaction. The two plasmon
branches correspond to in-phase (symmetric) and out-of-phase (anti-symmetric) charge-density
oscillations. When the difference between the radii is small in (a),
single-particle energies from the two shells almost coincide and
the symmetric and anti-symmetric plasmons lie close to each other and
are well above the regions where there exists Landau damping as shown
in Fig.\,\ref{FIG:2+}(a). This behavior is connected to a very weak
inter-shell Coulomb interaction which is scaled by $1-(R_1/R_2)^{2L+1}$ for $R_1\approx R_2$.
As the radius of the outer S2DEG is increased, one of the plasmon frequencies is pushed down.
Such an observation can be attributed to the strongly enhanced inter-shell Coulomb
interaction with $R_1\ll R_2$ as $L\gg 1$. When the outer radius is much larger than the
inner radius,  the lower plasmon branch is strongly Landau damped by particle-hole modes and its
intensity becomes very low.
\medskip

\begin{figure}[ht!]
\centering
\includegraphics[width=0.35\textwidth]{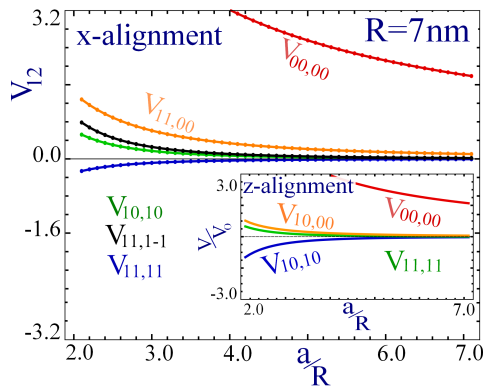}
\caption{(Color online) Coulomb matrix elements $\{V_{12}\}$ for two S2DEGs
of radius $R_1=R_2=R $\,nm in  units of $V_0=e^2/(4 \pi \epsilon_s R) = 2.98\, eV$ for $R=7\,nm$.
One sphere has its center at the origin while another has its center on the $x$-axis. The separation
between the centers of the two spheres is $a$ which is varied. The inset shows the dependence of the
Coulomb-interaction matrix elements when two spheres have their centers along the
$z$-axis for the same chosen parameters.}
\label{FIG:3}
\end{figure}

\noindent
Additionally, we have discovered that the Coulomb interaction between the shells is not simply
given by a power law but has oscillations due to the orbital motion of the S2DEG which
 is an interesting feature, that has not been discussed in the literature to our knowledge.

\begin{figure}[ht!]
\centering
\includegraphics[width=0.35\textwidth]{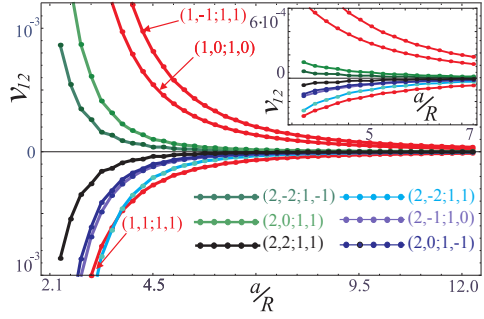}
\caption{(Color online) Comparison of the Coulomb matrix elements $\{V_{12}\}$ in the units
 $e^2/(  \epsilon_s R)$ with $  \epsilon_s=4 \pi \epsilon_0\epsilon_b$ for two S2DEGs when
$L=1$ and $L=2$. Here, one sphere has its center at the origin, while the other one
has its center on the $x$-axis. The radius of each S2DEG is
$R_1=R_2=R=1$\,nm. The separation between the centers of the spheres is $a$.
The inset shows the $a$ dependence within a smaller range.}
\label{FIG:7}
\end{figure}

In our calculations of the plasmon frequencies for the pair of
S2DEGs shown schematically in Fig.\,\ref{FIG:1}, we must truncate
the infinite matrix\,\cite{SSC-1996} in Eq.\,(\ref{eq1.51b}). Here, the
off-diagonal matrix elements involve the Coulomb interaction
$v_{1,\,2,\,3}=V_{LM,\,L^\prime M^\prime}(R_j,\,R_{j^\prime};\,a)$ (see Appendix\ A)
and their comparative values would determine whether a perturbation picture has   validity.
The origin of this Coulomb interaction comes from the optically-induced magnetic
dipoles by a finite photon angular momentum ($\ell\neq 0$) for anisotropic distribution of
electrons on the shell. If we choose the quantization axis of angular momentum along the probe
${\bf E}$-field direction, the   $L=0$ inter-sphere Coulomb interaction
has no contribution to the system. In the presence of a finite photon angular
momentum $L=1$, electron transition from the $\ell=0$ state to the $\ell=1$ state will occur.
The induced magnetic dipoles associated with the $\ell=1$ states of two displaced shells will couple
to each other either in phase or out of phase (split plasmon modes), leading to so-called Coulomb dimerization.
This leading magnetic coupling results from the action of  the magnetic field by the induced
oscillating electric dipole on one sphere on the induced magnetic dipole on another sphere.
Moreover, the coupling strength, which is scaled as $1/a$ for the far-field region
($a\gg R,\,\lambda_{\rm pl}$) or as $1/a^2$ for the near-field range ($a\gg R$ but $a\ll\lambda_{\rm pl}$),
is different when two shells are displaced along the direction either parallel to the
angular-momentum quantization axis (similar to $\pi$ bond for carbon atoms) or
perpendicular to the axis of quantization (similar to $\sigma$ bond). Here, $a$ is
the  separation between two displaced spheres, $R$ is the radius of spheres, $\lambda_
{\rm pl}$ is the plasmon wavelength and the coupling from acting of the magnetic field by the
induced oscillating magnetic dipole on one sphere on the induced magnetic dipole on another
sphere is vary small for large values of $a$. In this regard, we compare
the Coulomb matrix elements in Fig.\,\ref{FIG:3} when one sphere
is located at the origin while the other one has its center on the
$x$-axis at $(a,0,0)$ or along the $z$-axis at $(0,0,a)$. We chose
$R_1=R_2=R$, $L=L^\prime=1$ and $M,\,M^\prime=0,\,\pm 1$. There
are fewer non-zero Coulomb matrix elements when the spheres
are centered on the $z$-axis compared to when they are on the
$x$-axis. Additionally, the corresponding values for these
non-zero elements are not equal for the two orientations of the pair of S2DEGs,
which reflects the directionality in the plasmon-plasmon spatial correlation. There are
oscillations in $v_{1,\,2,\,3}$ in accordance with our semi-analytic results given for
large separations $a$ in Eqs.\,(\ref{semi-x}) and (\ref{semi-z})
in Appendix\ \ref{ap1} and the Coulomb interaction decreases with increasing
$a$. There is only one negative Coulomb matrix element for both configurations, implying
a weak bonding effect between two S2DEGs. Although the interaction $V_{00;\,00}(R_1,\,R_2;\,a)$
 between two electric dipole moments is always positive and a dominant one due to isotropic distribution
of electrons, it does not contribute to spherical plasmon excitations which require $L\geq 1$.
In Fig.\,\ref{FIG:7}, we also compare the Coulomb matrix elements with angular momentum quantum
numbers $L=1$ and $L=2$. These Coulomb matrix elements must be included if we would like to include
the coupling between these two higher angular momenta. Our results in
Fig.\ \ref{FIG:7},  however, show that to the lowest order, we may neglect these
couplings since they decay  fast with increasing separation $a$.
\medskip

\begin{figure}[ht!]
\centering
\includegraphics[width=0.35\textwidth]{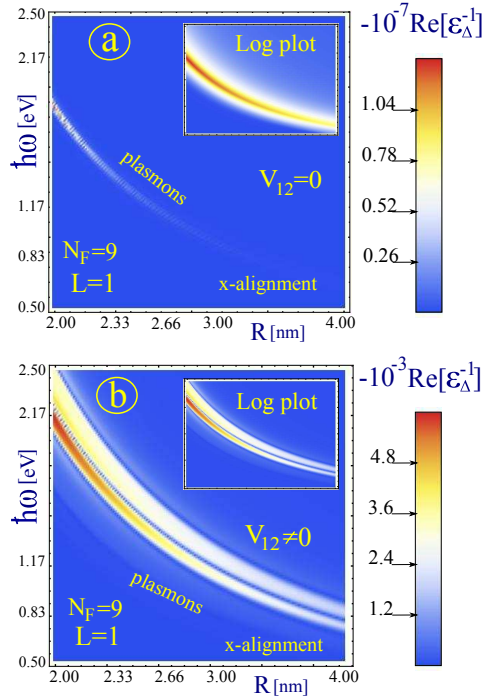}
\caption{(Color online) Comparison of the density plots
of frequency {\em vs.} radius $R$ for plasmon excitations when
$L=1$ and $N_F=9$ for a pair of coupled S2DEGs on the $x$-axis
with the inter-sphere Coulomb interaction included
($V_{12}\neq 0$) (lower panel) or excluded ($V_{12}= 0$) (upper panel).
The separation between the     spheres is $a-2R=0.1$\,nm.
Both insets show the logarithm of
one-plus the density obtained for each pair of values of
frequency and radius of the S2DEG of
the corresponding results.}
\label{FIG:5}
\end{figure}

\begin{figure}[ht!]
\centering
\includegraphics[width=0.27\textwidth]{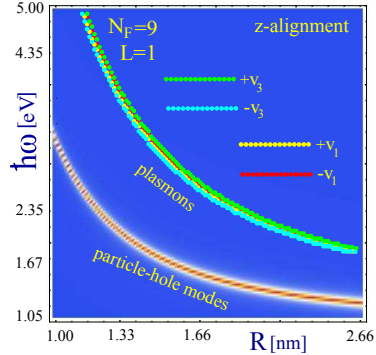}
\caption{(Color online) Density plots
of frequency {\em vs.} radius $R$ for plasmon  excitations when
$L=1$ and $N_F=9$ for a pair of coupled S2DEGs on the $z$-axis
with the inter-sphere Coulomb interaction included
($V_{12}\neq 0$). The two different non-zero potential matrix elements
are labeled $v_{1}$ and $v_{3}$  as defined in Appendix \ \ref{ap1}. The separation between
the  spheres is $a-2R-0.1$\,nm.}
\label{FIG:5+}
\end{figure}

In Figs.\ \ref{FIG:5} and \ref{FIG:5+}, we present results of
our calculations of the $L=1$ plasmon modes on a pair of
coupled S2DEGs with $R_1=R_2=R$ on the $x$-axis and $z$-axis, respectively. 
The difference between the two plots is striking but they still have some common
features. For example, the plasmon frequency is decreased as the
radius of the S2DEG is increased. For a chosen radius, the plasmon
frequency is slightly larger for the $z$-alignment. However, this small
difference in the plasmon excitation spectrum demonstrates that
{\em the plasma-plasma interaction is spatially correlated}. The inter-sphere Coulomb
interaction lifts the degeneracy of a plasmon mode on each sphere.
For $L=1$ and $M=0,\,\pm 1$, these two sets of modes are coupled to form  three in-phase
symmetric and three out-of-phase antisymmetric modes of charge-density oscillations.
Therefore, one expects that the plasma mode equation
would in general yield six solutions. However, some plasmon frequencies
are degenerate while others may be close. We emphasize
that the semi-analytic forms of the Coulomb interactions in
Eqs.\,(\ref{semi-x}) and (\ref{semi-z}) do not scale as
a point-like dipole-dipole coupling at large separation. The angular momentum
quantum numbers $L$, $M$ also determine these Coulomb matrix elements, i.e., the 
plasmon-plasmon interaction depends on the spatial profile of an incident light beam.
This implies that the bonding process in the Coulomb dimerization is directional, depending on the
angular distribution of electrons for $\ell\neq 0$, similar to $\sigma$ and $\pi$ bonds 
between two carbon atoms. Figure\ \ref{FIG:5}(b) shows that when the inter-sphere Coulomb
interaction is included, the intensity of the density plots  for plasmon excitations is  
enhanced from their values in Fig.\,\ref{FIG:5}(a), indicating a dimerization process 
between two spatially separated S2DEGs. The reason we truncated the matrix in our calculations 
was to see the effect of the Coulomb coupling  between the shells on the lowest plasmon modes. Each of these
modes has different intensity arising from the value of the loss function. By including
the matrix elements which involve the $L=2$ angular momentum, there will be additional plasmon
modes, which will not affect these six lowest modest substantially if the two shells are not 
too close to each other (see Fig.\,\ref{FIG:7}). These results in conjunction with those
in Figs.\,\ref{FIG:5}(b) and \ref{FIG:5+} clearly demonstrate
the existence of plasmon-coupling based dimerization, as well as its significance, in this system.
\medskip

\begin{figure}[ht!]
\centering
\includegraphics[width=0.35\textwidth]{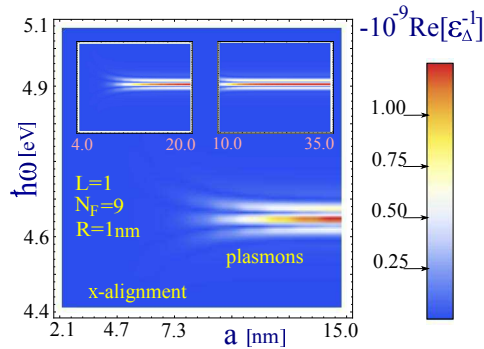}
\caption{(Color online) Density plot
of frequency {\em vs.} separation $a$ for plasmon
excitations when $L=1$ for a pair of
S2DEGs each of radius $R=1$\,nm. One sphere is at the origin and
the other is centered on the $x$-axis. The inset shows how the
plasmon excitations merge to form a single branch as $a\gg R$.
Here, we have fixed $N_F=9$.}
\label{FIG:6}
\end{figure}

Finally, we investigated in Fig.\,\ref{FIG:6} the dependence of
plasma frequency on separation $a$ between two S2DEGs with their centers on the
$x$-axis. As expected, the split plasmon branches merge into a
single branch for sufficiently large separations, as depicted in the inset of
Fig.\,\ref{FIG:6}. As we mentioned above, the multiplicity of plasmon modes
could be less than six distinct solutions which is determined by the strengths
of the inter-sphere Coulomb matrix elements.  The calculated $1/a$ dependence of 
the inter-sphere Coulomb interaction directly verifies the magnetic-field coupling 
mechanism for $a\gg R,\,\lambda_{\rm pl}$ in the dimerization process.

\section{Concluding   Remarks}
\label{sec4}

In this paper, we presented a formalism for calculating
plasma excitations for a pair of Coulomb coupled spherical
electron gases. The RPA was used in  this investigation.
The S2DEG is a first-order approximation for calculating
electronic properties of fullerenes when the lattice
structure, band width from electron hopping and radial motion may be neglected.
However, we may incorporate a more realistic energy band structure into the
polarization function through a form factor by making use of the results
presented in Ref.\,[\,\onlinecite{Lett2}]. This would also account for the prescribed number of
electrons on the fullerene. The plasmon excitation formula for a pair
of Coulomb coupled S2DEGs is different when the shells are concentric compared
to when these shells are side-by-side. For one thing, the angular
momentum is a good quantum number for concentric shells, but all
angular momenta on two displaced shells are coupled. In the latter case, we calculated
the plasmon modes approximately to lowest order by including only
the dominant Coulomb matrix elements which were obtained analytically.
Additionally, we have demonstrated that the frequency of plasmon
excitation for a pair of displaced and Coulomb coupled S2DEGs depends on both
the separation between their centers as well as whether their
centers lie along the axis of quantization or not. This is a
consequence of the functional dependence  of the Coulomb
matrix elements on spatial orientation.

\par
\medskip

We note that  spectral correlations  have been observed experimentally for metallic nanoparticles
\,\cite{Yang1} In that work,  the plasmons for pairs  were studied using polarization-selective 
total internal reflection.  Their measurements show that the frequencies for the coupled plasmon 
modes  depend on whether the incident light wave vector perpendicular and parallel to the dimer 
axis. Related work on dimer plasmons has been conducted by Nordlander, et al. (Ref.[\onlinecite{Nord1}])
with the conclusion  that the hybridized plasmon energy arising from individual metallic nanoparticles 
is determined by the  orientation of the inter-particle axis with respect to the axis of polarization 
of the two constituents modeled as incompressible spherical liquids. Although  our model differs from that
in Ref.[\onlinecite{Nord1,Prodan}] the conclusions about the existence of anisotropy in the plasmon excitation
energies in these systems are in agreement. Similar effects are also expected to be observed in the case of 
nano-eggs: non-concentric multishells of nanoparticles. The hybridization of the plasmons has been proven to 
be an adequate and precise method to describe the plasmonic structure \cite{Wu,Bardhan}. The field enhancement, 
corresponding to the resonant excitation of plasmons, was reported to be much larger in the case of concentric 
nanoparticles, which support indirectly the concept of plasmon spatial correlation.

\par
\medskip

Generally, the angular momentum of light may be carried by either orbital motion (helicity) 
or spin motion (circular polarization). When the incident light has zero angular momentum 
with $L=0$, the electric field generated from the induced isotropic electric dipole moment for 
the $\ell=0$ electron state  on one sphere may couple to an induced electric dipole moment 
for other $\ell=0$ electron states on another displaced sphere. However, such an isotropic plasmon excitation
is associated with a change of the radial quantum number, and is prohibited in our model 
for spherical shells of electron gases. If a finite angular momentum of light with $L=1$ is 
used for incidence, on the other hand, the magnetic field generated from the induced oscillating
electric dipole moment on one sphere can couple to the induced magnetic dipole moment on 
another displaced sphere. This unique inter-sphere magnetic (plasmon) coupling, which is
associated with the magnetic dipole moment for $\ell=1$, becomes anisotropic in space, depending
on the displacement of two spheres parallel or perpendicular to the direction of a probe electric field.
In addition, such an oscillating electric dipole moment based inter-sphere magnetic coupling 
directly leads to dimerization of electron gases on two spheres. For the plasmon excitation 
with $L\geq 2$, the higher angular-momentum component of specific incident light beam is 
required, such as  a helical or a Bessel light beam. In principle, the effect of plasmon 
coupling predicted in this paper should be experimentally observable by using   light with a 
finite angular momentum for incidence and rotating the sample by $90^{\rm o}$
for showing its directional bonding effect.
The key feature presented in  this paper is the broken rotational symmetry by coupling 
between two center-displaced S2DEGs based on photo-excited electron density fluctuations. 
Here, the quantization axis of the system is selected by the probe electric field. Such a 
directional plasmon-correlation effect will be lost if two S2DEGs are projected onto a plane,
i.e., a pair of quantum rings\,\cite{huang1}, because the quantization axis is always 
perpendicular to the plane of the rings. For a S2DEG, we obtain a degeneracy in single-electron 
kinetic energies with respect to $m$ (angular momentum number along the axis of quantization). 
But, this degeneracy is reduced to $\pm m$ for a quantum ring. In order to completely remove 
the $m$-degeneracy, an external magnetic field ${\bf B}$ can be applied to the system.
In this case, a strong magnetic field will change the S2DEG simply to a Landau quantized 
S2DEG\,\cite{huang2} with kinetic energy $\sim\hbar\omega_c$
but change a quantum ring to a classical point mass rotating around a circle with
reduced inertia and   angular velocity $\omega_c$, where $\omega_c=eB/\mu^\ast$.
The study of spatial correlation of magneto-S2DEGs on two displaced spheres is under 
investigation.

\begin{acknowledgments}
This research was supported by  contract \# FA 9453-07-C-0207 of
AFRL. DH would like to thank the support from the Air Force Office of Scientific Research (AFOSR).
\end{acknowledgments}


\appendix

\section{Calculations of Coulomb interaction matrix elements}
\label{ap1}

In Appendix\ \ref{ap1}, we demonstrate how we obtained semi-analytic expressions for the 
potential matrix elements $V_{LM,\,L^\prime M^\prime}(R_1,\,R_2;\,a)$ with different 
eigenstates labeled by $L,M$ and $L^\prime,M^\prime$,
corresponding to each sphere. Our primary consideration arises
when the two shells are centered on the $x-$axis, i.e., an axis
perpendicular to the axis of the angular momentum quantization.
A simple case when the two spheres are centered on the $z-$axis
will be briefly discussed at the end of Appendix\ \ref{ap1}.

\subsection{Plane-Wave Expansion Method}

We now describe how each of the matrix elements could
be expanded as a linear combination of triple spherical
Bessel function integrals. In this regard, we must evaluate the
following integral

\begin{equation}
{\cal I}\equiv\int \frac{d^3{\bf q}}{q^2}\,j_L(qR_1)j_{L^\prime}(q R_2)\,
Y_{L,M}^{\ast}(\Omega_{\hat{\bf q}})Y_{L^\prime,M^\prime}(\Omega_{\hat{\bf q}})\,
\label{aa1}
\texttt{e}^{-iq_xa}\ .
\end{equation}
First, we present in spherical coordinates  $\Omega_{\hat{\bf q}}=\{1,\,\theta,\,\phi\}$
and $\hat{\bf e}_x= \{1,\,\theta=\frac{\pi}{2},\,\phi=\pi\}$
(\underline{Note}: $\phi=$ either $0$ or $\pi$ depending on whether we need to calculate 
$\texttt{e}^{iq_xa}$  or $\texttt{e}^{-iq_xa}$).
The standard plane-wave expansion over spherical harmonics gives

\begin{equation}
\texttt{e}^{-iq_xa}=4\pi\sum\limits_{\lambda}\sum\limits_{\mu=-\lambda}^{\lambda}\,
(i)^{\lambda}\,j_{\lambda}(qa)\,Y_{\lambda,\mu}^\ast(\Omega_{\hat{\bf q}})Y_{\lambda,\mu}(\Omega_{\hat{\bf e}_x})\ .
\label{plane}
\end{equation}
By making use of this result, the integral in Eq.\,(\ref{aa1}) turns into

\[
{\cal I}=(4\pi)\sum\limits_{\lambda,\,\mu}\,\int\limits_{0}^{\infty} dq\,(i)^{\lambda}\,j_L(q R_1)j_{L^\prime}(q R_2)
j_\lambda(q a)\,Y_{\lambda,\mu}(\Omega_{\hat{\bf e}_x})
\]
\begin{equation}
\times\int d\Omega_{\hat{\bf q}}\,Y_{L,M}^{\ast}(\Omega_{\hat{\bf q}})Y_{L^\prime,M^\prime}(\Omega_{\hat{\bf q}})
Y_{\lambda,\mu}^{\ast}(\Omega_{\hat{\bf q}})\ .
\end{equation}
By recalling the "triple-Y" integration formula

\begin{eqnarray}
&&\int d\Omega_{\hat{\bf q}}\,Y_{L,M}(\Omega_{\hat{\bf q}})
Y_{L^\prime,M^\prime}(\Omega_{\hat{\bf q}})Y_{\lambda,\mu}(\Omega_{\hat{\bf q}})
\nonumber\\
&=& \sqrt{\frac{(2L+1)(2L^\prime+1)(2 \lambda+1)}{4\pi}}
\left({ \begin{array}{ccc}
L & L^\prime  & \lambda\\ 0 & 0 & 0  \end{array} }\right)
\left({ \begin{array}{ccc} L & L^\prime  & \lambda\\
M & M^\prime & \mu  \end{array} }\right)
\end{eqnarray}
and the identity $Y_{L,M}(\theta,\,\phi)=(-1)^M\,Y^{\ast}_{L,-M}(\theta,\,\phi)$, we 
finally obtain from Eq.\,(\ref{eq1.51})

\begin{eqnarray}
V_{LM,\,L^\prime M^\prime}(R_1,\,R_2;\,a)&=&\frac{8e^2}{\epsilon_s}
\sum\limits_{\lambda,\,\mu}\,(-1)^{\lambda/2-M-\mu}\,Y_{\lambda,\mu}(\Omega_{\hat{\bf e}_x})
\nonumber\\
&\times & \sqrt{\frac{(2L+1)(2L^\prime+1)(2\lambda+1)}{4\pi}}
\left({\begin{array}{ccc} L & L^\prime  & \lambda\\ 0 & 0 & 0  \end{array} }\right) \left({ \begin{array}{ccc}
L & L^\prime  & \lambda\\ M & M^\prime & \mu  \end{array} }\right)
\nonumber\\
&\times &\int\limits_{0}^{\infty} dq\,j_L(qR_1) j_{L^\prime}(qR_2)j_\lambda(qa)\ .
\end{eqnarray}
Since there are only a few non-zero terms in that summation
\textit{(see next section of Appendix\ \ref{ap1})}, we simply write

\begin{equation}
V_{LM,\,L^\prime M^\prime}(R_1,\,R_2;\,a)=\sum\limits_{\lambda}\,
\mathcal{C}_{\lambda}(L,M;\,L^\prime,M^\prime)\int\limits_{0}^{\infty} dq\,
j_L(qR_1)j_{L^\prime}(qR_2)j_\lambda(qa)\ .
\end{equation}
For $z \to  \infty$, we have

\begin{equation}
j_L(z)\approx \frac{1}{z}\,\sin\left({z-\frac{\pi L}{2}}\right)\ ,
\end{equation}
so that we obtain, in the limit $a\to\infty$,

\[
V_{LM,\,L^\prime M^\prime}(R_1,\,R_2;\,a)\approx\frac{1}{a}\sum\limits_{\lambda}
\mathcal{C}_{\lambda}(L,M;\,L^\prime,M^\prime)
\]
\begin{equation}
\times\int\limits_0^\infty\,\frac{dq}{q}\,j_L(qR_1) j_{L^\prime}(q R_2)
\sin\left(qa-\frac{\pi\lambda}{2}\right)\ ,
\label{semi-x}
\end{equation}
which shows that  the asymptotic behavior of Coulomb
interaction exhibits oscillations with respect to sphere separation $a$.

\subsection{Analytic Evaluation of the Angular Integrals for Potential Matrix Elements ($x-$ alignment)}
\label{11}

One may verify that it is possible to perform the angular $\phi$ 
and $\theta$  integrations analytically for all potential matrix
elements in a relatively straightforward way. The starting point is to
evaluate a three-dimensional integral in spherical coordinates:

\begin{eqnarray}
&&V_{LM,\,L^\prime M^\prime}(R_1,\,R_2;\,a)=\frac{2e^2}{\pi\epsilon_s}
\int \frac{d^3{\bf q}}{q^2}\,j_L(R_1q)j_{L^\prime}(R_2q)\,
Y_{L,M}^\ast(\theta,\,\phi)Y_{L^\prime,M^\prime}(\theta,\,\phi)\,\texttt{e}^{-iq_xa}
\nonumber \\
&=&\frac{2e^2}{\pi\epsilon_s}\int\limits_0^\infty dq\,j_L(R_1q)j_{L^\prime}(R_2q)\int\limits_0^\pi \sin\theta\,d\theta\int\limits_0^{2\pi}
d\phi\,Y_{L,M}^\ast(\theta,\,\phi) Y_{L^\prime,M^\prime}(\theta,\,\phi)\,\texttt{e}^{-iq_xa} \ .
\end{eqnarray}
We perform the $\phi-$integration first and label the result as $\mathcal{I}_\phi$.
After the $\theta$-integration is completed, the final angular integral will be referred
to as $\mathcal{I}_\theta$.
\medskip

In spherical coordinates $q_x=q\,\sin\theta\cos\phi$, therefore, one writes

\begin{equation}
\label{phi1}
\mathcal{I}_\phi(q,\,\theta)=\int\limits_0^{2\pi} d\phi\,
\texttt{e}^{-i\,aq\sin\theta\cos\phi}\ .
\end{equation}
In order to obtain a closed-form analytic result from Eq.\,(\ref{phi1}),
we use the Jacobi-Anger identity

\begin{equation}
\label{JA}
\texttt{e}^{i\xi\cos\phi}=\sum\limits_{m=-\infty}^\infty\,i^m\,J_m(\xi)\,\texttt{e}^{im\phi}\ ,
\end{equation}
where $J_m(x)$ stands for Bessel functions of the first kind.

\medskip
It follows from Eq.\,(\ref{JA}) and the exponential
$\phi$-dependence of spherical harmonics $Y_{L,\,\pm M}(\theta,\,\phi)$,
that the order of the only remaining non-zero term contains the
\textit{Bessel function}, determined by the difference between  $M$
and $M^\prime$ values, namely by $\vert M-M^\prime\vert$. Consequently,
we classify all the results of $\phi$-integration by $\vert M-M^\prime\vert$.
For our present consideration with $L,\,L^\prime=1$ this difference could only be $0,\,1$
or $2$.
\medskip

Since the only way to obtain an imaginary result  for $\mathcal{I}_\phi(q,\,\theta)$
in Eq.\eqref{plane} comes from  $i^m$ (see Eq.\eqref{exp}), we see that

\begin{equation}
\mathcal{I}_\phi(q,\theta) \backsim i^{\vert M- M' \vert}\ .
\end{equation}
Therefore, the result of $\phi$-integration (and, consequently, the potential
matrix element) will be \textbf{imaginary} if $\vert M-M^\prime\vert=1$. For the
relevant case $L=L^\prime=1$, all the elements with $\vert M - M^\prime \vert =1$,
i.e., $M=\pm 1$, $M^\prime=0$, or vice versa, are equal to zero due to a specific
symmetry in the $\theta$-integration (check the matrix elements in Eq.\,(\ref{matgeneral})),
so that all the potential matrix elements are \textit{real}.
\medskip

\subsection{Summary of Relevant Potential Matrix Elements for $L=1$}

As we are now going to obtain, the modifications $\texttt{e}^{-i qa\sin\theta\cos\phi}
\Rightarrow\texttt{e}^{iq\sin\theta\cos\phi}$, $\texttt{e}^{i\phi}\Rightarrow \texttt{e}^{-i\phi}$
and $\texttt{e}^{2i\phi} \Rightarrow \texttt{e}^{-2i\phi}$ do not alter the values of all real
$\phi$-integrals. As long as only these elements result in a non-zero $\theta$-integral,
the potential sub-matrices $A$ and $B$ in Eq.\,(\ref{blockm}) are identical.
\medskip

We noted in Sec.\,\ref{11} that the result of the $\phi$-integration is determined by 
$ \vert M-M^\prime \vert $ and does not depend on each individual $M,\,M^\prime$ value. 
Consequently, there are only \textit{three different non-zero potential matrix elements}, which 
will be later referred to as $v_{1,\,2,\,3}$, respectively.
\medskip

Let us now briefly provide the integration results for each non-zero potential matrix element
$V_{L\, M,\,L^\prime \, M^\prime}(R_1,\,R_2;\,a)$. As mentioned above, $L=L^{\prime}=1$ for
all cases.

\subsubsection{$M=M^\prime=0$}

\begin{align}
& \mathcal{I}_\phi(q,\,\theta) =  \int\limits_0^{2\pi} d\phi\,
\texttt{e}^{-iqa\sin\theta\cos\phi} \left|Y_{1,\,0}(\theta, \phi)\right|^2
=\frac{3}{2} \cos^2\theta \,J_0(aq\sin\theta)\ ,
\nonumber\\
& \mathcal{I}_{\theta}(q) =\frac{3}{2} \int\limits_0^\pi\,J_0(aq
\sin\theta)\,\sin\theta\cos^2\theta\,d\theta=3\,\frac{\sin (aq)-aq\,\cos (aq)}{(aq)^3}\ ,
\nonumber \\
& v_3\equiv V_{10,\,10}(R_1,\,R_2;\,a) = \frac{6 e^2}{ \pi \epsilon_s}\int\limits_0^{\infty}\,dq\,
\frac{\sin (aq) -aq\,\cos(aq)}{(aq)^3}\,j_1(R_1q)j_1(R_2 q)\ .
\end{align}
Here we write the final answer, taking into account the coefficient $2 e^2/\pi \epsilon_s$ 
for all potential matrix elements.

\subsubsection{ $M=M^\prime \neq 0$}

\begin{align}
\label{r2}
& \mathcal{I}_\phi(q,\,\theta) = \int\limits_0^{2\pi} \texttt{e}^{-iqa\
sin\theta\cos\phi}\,\vert Y_{1,\,{\pm 1}} (\theta, \phi) \vert^2 \, d\phi=\frac{3}{4} 
\sin^2\theta  J_0(aq\,\sin\theta)\ ,
\nonumber\\
& \mathcal{I}_{\theta}(q) =\frac{3}{4} \int\limits_0^\pi J_0(aq\,\sin\theta)
\,\sin^3\theta\,d\theta= \frac{3}{2}
\frac{ aq\,\cos(aq)-\left({(aq)^2-1}\right)\,\sin(aq)}{(aq)^3}\ ,
\nonumber\\
&  v_1 \equiv V_{1\,-1,\,1\,-1}(R_1,\,R_2;\,a)=V_{1\,1,\,1\,1}(R_1,\,R_2;\,a)= \frac{3 e^2}{\pi \epsilon_s}
\int\limits_0^{\infty} dq\,\frac{ aq\,\cos(aq)+\left({(aq)^2-1}\right)\,\sin(aq)}{(aq)^3}\,
j_1(R_1q)j_1(R_2q)\ .
\end{align}

\subsubsection{$\vert M-M^\prime\vert=1 \Rightarrow  M \lor M^\prime = 0$}

\begin{align}
& \mathcal{I}_\phi(q,\,\theta) \backsimeq \int\limits_0^{2\pi} d\phi\,
\texttt{e}^{-iqa\sin\theta\cos\phi}\,\ex^{\pm i\phi}= 2 \pi i\,J_1(aq\,\sin\theta)\ ,
\nonumber \\
& \mathcal{I}_{\theta}(q) \backsimeq \int\limits_0^\pi\,J_1(aq\,\sin\theta)\,
\sin^2\theta\cos\theta\,d\theta= 0\ .
\end{align}
Therefore, we confirm our previous finding, that all potential matrix elements with one 
$M =\pm 1$, $M^{\prime}=0$ and vice versa \emph{are equal to zero} (\textit{four} elements in 
each sub-matrix).

\subsubsection{$\vert M-M^\prime \vert  = 2 \Rightarrow M=-M^\prime$}

\begin{align}
& \mathcal{I}_\phi(q,\,\theta) = \int\limits_0^{2\pi} d\phi\, \left[{Y_{1,\,
{\pm 1}}(\theta, \phi)}\right]^{\ast} Y_{1,\,{\mp 1}}(\theta, \phi) \texttt{e}^{-iqa
\sin\theta\cos\phi}= \frac{3}{4} \sin\theta^2 \,J_2(aq\,\sin\theta)\ ,
\nonumber\\
& \mathcal{I}_{\theta}(q) =\frac{3}{4} \int\limits_0^\pi\, J_2(aq\sin\theta)\,\sin^3
\theta\,d\theta= -\frac{3}{2} \frac{3 aq\,\cos (aq)-\left({(aq)^2-1}\right)\,\sin (aq)}{(aq)^3}\ ,
\nonumber\\
& v_2 \equiv V_{1\,-1,\,1\,1}(R_1,\,R_2;\,a)=V_{1\,1,\,1\,-1}(R_1,\,R_2;\,a)
=- \frac{3e^2}{\pi \epsilon_s}\int\limits_0^{\infty} dq\,\frac{3 aq\,\cos (aq)-
\left({(aq)^2-3}\right)\,\sin (aq)}{(aq)^3}\,j_1(R_1q)j_1(R_2q)\ .
\end{align}

\subsection{$z$-alignment}

We now turn to a brief discussion of the  case when the two
spheres have their centers on the  $z$-axis, which is also
the axis of the angular momentum quantization. Significant
simplification comes from the fact that the exponential term
$\texttt{e}^{-i q_z a}=\texttt{e}^{-i q a \cos\theta}$ which does not depend
on the azimuthal angle $\phi$. Consequently, the $\phi$-dependence of each
potential matrix element is now determined solely by the exponential part or phase
of the spherical harmonics. Since

\begin{equation}
\int_0^{2\pi}\,d\phi\,\texttt{e}^{i(M-M^\prime)\phi}
=2\pi\,\delta_{MM^\prime} \ ,
\label{exp}
\end{equation}
we can conclude that
\medskip

\noindent
(a) only elements with $M=M^\prime$ are non-zero;
\par

\noindent
(b) $V_{1\,-1,\,1\,-1}(R_1,\,R_2;\,a)=V_{1\,1,\,1\,1}(R_1,\,R_2;\,a)$.
\medskip

\noindent
Consequently, we need to evaluate only two  non-zero elements, contributing to the
plasmon equations, namely $V_{1-1,\,1\,-1}(R_1,\,R_2;\,a)=V_{1\,1,\,1\,1}(R_1,\,R_2;\,a)$
with $M=M^\prime \neq 0$ as well  $V_{10,\,10}(R_1,\,R_2;\,a)$. The calculation is now
straightforward:

\begin{align}
& v_1 \equiv V_{1\,1,\,1\,1}(R_1,\,R_2;\,a) = \int\limits_0^{\infty} dq \
j_1(R_1 q) j_1(R_2 q) \int\limits_0^{\pi} \texttt{e}^{- i q a \cos\theta} 
\sin\theta d\theta \int\limits_0^{2\pi} d\phi \, \vert Y_{1,\,1}(\theta; \phi) \vert^2  =
\nonumber\\
& = \frac{6 e^2}{\pi \epsilon_s}\int\limits_0^{\infty}\,dq\,\frac{\sin (aq) -aq\,\cos (aq)}
{(aq)^3}\,j_1(R_1q)j_1(R_2q)\ .
\label{v1}
\end{align}
Here, again, we provide the final answer, taking into account the coefficient 
$2 e^2/\pi \epsilon_s$ for all potential matrix elements.
\medskip

Correspondingly, the remaining potential $V_{10,\,10}(R_1,\,R_2;\,a)$ is as follows:

\begin{align}
& v_3 \equiv V_{1\,0,\,1\,0}(R_1,\,R_2;\,a) = \int\limits_0^{\infty} dq j_1(R_1 q) j_1(R_2 q) 
\int\limits_0^{\pi} d\theta\ \texttt{e}^{- i q a \cos\theta} \sin\theta   \int\limits_0^{2\pi} d\phi \ 
\left({ Y_{1,\,0}(\theta; \phi) }\right)^2   
\nonumber\\
& = \frac{6 e^2}{\pi \epsilon_s}\int\limits_0^{\infty}\,dq\,\frac{\left({(aq)^2 - 2}\right)
 \sin(aq) + 2 a q \cos(aq)}{(aq)^3}\,j_1(R_1q)j_1(R_2q)\ .
\label{semi-z}
\end{align}
We assign $v_3$ to the second non-zero potential matrix element to keep the notations 
uniform with the previous section.
\medskip

Finalizing this Appendix section, we provide the expression for $V_{0\,0,\,0\,0}(R_1,\,R_2;\,a)$, 
representing the highest values of the interaction potential. This element, however, does not 
provide any contribution to the plasmon equations, since the polarization function is zero for L=0:

\begin{equation}
V_{0\,0,\,0\,0}(R_1,\,R_2;\,a) =\frac{3 e^2}{\pi \epsilon_s}\int\limits_0^{\infty}\,dq\,
 j_0(a q) j_0(R_1q)j_0(R_2q)  \ .
\label{v00}
\end{equation}
One can easily verify exactly that the same expression could be obtained for $V_{0\,0,\,0\,0}(R_1,\,R_2;\,a)$ 
in the case of two $x-$aligned spherical shells, which could be explained by the fact the interaction between the shells with $L=0=L^\prime=0$ is obviously spherically-symmetric.

\section{Matrix Transformations and the Calculations of Determinants }
\label{ap2}

\subsection{$x$-aligned shells}

For us to simplify the evaluation of the determinant, we use
a property of the determinant of a block matrix, i.e.,
for such a matrix

\begin{equation}
{\cal \underline{M}}_B=\left[{  \begin{array}{cc} {\cal \underline{D}}_1 & {\cal \underline{A}} \\
                               {\cal \underline{B}} & {\cal \underline{D}}_2  \end{array}  }\right]\ ,
\label{blockm}
\end{equation}
its determinant is given by
$\mbox{Det}\left[{\cal \underline{M}}_B\right]=\mbox{Det}\left[{\cal \underline{D}}_1\otimes{\cal \underline{D}}_2
-{\cal \underline{A}}\otimes{\cal \underline{B}}\right]$.
For all separations $a>R_{1}+R_{2}$, the Coulomb matrix
elements in the diagonal blocks ${\cal \underline{D}}_1$
and ${\cal \underline{D}}_2$ are larger than those in the off-diagonal blocks
${\cal \underline{A}}$ and ${\cal \underline{B}}$.
\medskip

We now write explicitly

\begin{equation}
{\cal \underline{D}}_1=\left[{  \begin{array}{ccc}
\epsilon_{L=1}(R_1,\,\omega) & 0 & 0 \\
0 & \epsilon_{L=1}(R_1,\,\omega) & 0\\
0 & 0 & \epsilon_{L=1}(R_1,\,\omega)\end{array} }\right]\ ,
\end{equation}
and

\begin{equation}
{\cal \underline{D}}_2=\left[{  \begin{array}{ccc}
\epsilon_{L=1}(R_2,\,\omega) & 0 & 0 \\
0 & \epsilon_{L=1}(R_2,\,\omega) & 0\\
0 & 0 & \epsilon_{L=1}(R_2,\,\omega)\end{array}  }\right]\ .
\end{equation}
Let us consider identical spheres, i.e., equal radius $R_1=R_2=R$
and chemical potential  $\mu_1=\mu_2$. In such a case, we have

\begin{equation}
{\mbox Det}\left[{\cal D}_1\right]={\mbox Det}\left[{\cal D}_2\right]=
\left[\epsilon_{L=1}(R,\,\omega)\right]^3 = \left[1-\frac{e^2}{3\epsilon_sR}\,\Pi_{L=1}(\omega)\right]^3\ ,
\end{equation}
and, correspondingly, since both blocks ${\cal D}_1$ and ${\cal D}_2$ are diagonal,
we obtain

\begin{equation}
{\mbox Det}\left[{\cal \underline{D}}_1\otimes{\cal \underline{D}}_2\right] = \left[\epsilon_{L=1}(R,\,\omega)\right]^6 = \left[1-\frac{e^2}{3\epsilon_sR}\,\Pi_{L=1}(\omega)\right]^6\ .
\end{equation}
From these results, we see that a density plot of the imaginary part
of $1/\mbox{Det}\left[{\cal \underline{M}}_B\right] $ will show that the frequencies of the
particle-hole excitations and the plasmons will appear almost
as a power-law dependence on the imaginary part of
$1/\left[\epsilon_{L=1}(R,\,\omega)\right]^6$, with a correction due to the
inter-sphere Coulomb interaction.
\medskip

Previously, we obtained the potential matrices in Eq.\,(\ref{eq1.51b}) with given $v_{1,\,2}$ by

\begin{equation}
\label{matgeneral}
{\cal \underline{A}}={\cal \underline{B}}=\Pi_{L=1}(R,\,\omega)
\left[{  \begin{array}{ccc}
v_1  &  0  & v_2\\
0  &  v_3  &  0\\
v_2  &  0  & v_1
\end{array}  }\right]\ .
\end{equation}
Therefore, ${\mbox Det}\left[{\cal \underline{A}}\right]={\mbox Det}\left[{\cal \underline{B}}\right]=(v_1-v_2)(v_1+v_2)\,v_3$ and their product
${\cal \underline{A}}\otimes{\cal \underline{B}}$
for $R_1=R_2=R$ yields the off-diagonal corrections in the following form

\begin{equation}
{\cal \underline{A}}\otimes{\cal \underline{B}}=
\left[\Pi_{L=1}(R,\,\omega)\right]^2\left[{  \begin{array}{ccc}
(v_1)^2+(v_2)^2  &  0  & 2v_1v_2\\
0  &  (v_3)^2  &   0\\
2v_1v_2   &   0   & (v_1)^2+(v_2)^2
\end{array}  }\right]\ .
\end{equation}
\medskip

Finally, the complete determinant, which yields the  electronic
excitations, can expressed as

\begin{gather}
\notag
Det\left[{\cal \underline{D}}_1\otimes{\cal \underline{D}}_2-{\cal \underline{A}}\otimes{\cal \underline{B}}\right]=
Det\left[{     \begin{array}{ccc}
d_1  &   0   &   -2{\cal Q}_1\,v_1v_2\\
0   &   d_2   &   0\\
-2{\cal Q}_1\,v_1v_2   &   0    &   d_2
\end{array}   }\right]\ ,
\end{gather}
where

\begin{eqnarray}
d_1 &=& \epsilon_{L=1}(R_1,\,\omega)\,\epsilon_{L=1}(R_2,\,\omega)-{\cal Q}_1\left[(v_1)^2+(v_2)^2\right]\ ,\\
d_2 &=& \epsilon_{L=1}(R_1,\,\omega)\,\epsilon_{L=1}(R_2,\,\omega)-{\cal Q}_1\,(v_3)^2\ ,\\
{\cal Q}_1 &=& \frac{2e^2}{\pi\epsilon_s}\,\Pi_{L=1}(R_1,\,\omega)\,\Pi_{L=1}(R_2,\,\omega)\ .
\end{eqnarray}

\subsection{$z$-aligned shells}

From the previous discussion it follows that, for the $q_z$-case, each potential
sub-matrix ${\cal \underline{A}}$ and ${\cal \underline{B}}$ is diagonal and has the following form:

\begin{equation}
\label{matgeneralz}
{\cal \underline{A}}={\cal \underline{B}}=   \left[{  \begin{array}{ccc}
v_1  &   0   &   0\\
0   &   v_3   &   0\\
0   &   0   &   v_1
\end{array}  }\right]
\end{equation}
with $v_1$ and $v_3$ given by Eqs.\,(\ref{v1}) and (\ref{semi-z}).
The reduced $3\times 3$ matrix is

\begin{equation}
\footnotesize
\notag
\left[{ \begin{array}{ccc}
\epsilon_{L=1}(R_1,\omega)\,\epsilon_{L=1}(R_2,\,\omega)-{\cal Q}_2\,v_1^2  &  0  &  0\\
0  &  \epsilon_{L=1}(R_1,\,\omega)\,\epsilon_{L=1}(R_2,\,\omega)-{\cal Q}_2\,v_3^2  &  0\\
0  &  0  &\epsilon_{L=1}(R_1,\,\omega)\,\epsilon_{L=1}(R_2,\,\omega)-{\cal Q}_2\,v_1^2
\end{array}   }\right]\ ,
\end{equation}
where ${\cal Q}_2=\Pi_{L=1}(R_1,\,\omega)\,\Pi_{L=1}(R_2,\,\omega)$
and $\Pi_{L}(R_j,\,\omega)$ is the polarization function for
the shell of radius $R_j$ given in Eq.\,(\ref{gae13}).
This leads to the following equations for plasmon modes

\begin{eqnarray}
&&\epsilon_{L=1}(R_1,\,\omega)\,\epsilon_{L=1}(R_2,\,\omega)\pm
\left[\Pi_{L=1}(R_1,\,\omega)\,\Pi_{L=1}(R_2,\,\omega)\right]^{1/2}\,v_1 = 0\ ,
\nonumber\\
&&\epsilon_{L=1}(R_1,\,\omega)\,\epsilon_{L=1}(R_2,\,\omega)\pm
\left[\Pi_{L=1}(R_1,\,\omega)\,\Pi_{L=1}(R_2,\,\omega)\right]^{1/2}\,v_3 = 0\ .
\end{eqnarray}

\end{document}